\documentclass[reprint,superscriptaddress,showpacs,showkeys,amsmath,amssymb,aps,nofootinbib,preprintnumbers]{revtex4-1}

\usepackage{graphicx}
\usepackage[mathlines,modulo]{lineno}
\usepackage[english]{babel}
\usepackage{color}
\usepackage{ulem}

\def \nutau {$\nu_\tau~$}

\def\lapproxeq{\lower .7ex\hbox{$\;\stackrel{\textstyle <}{\sim}\;$}}
\def\gapproxeq{\lower .7ex\hbox{$\;\stackrel{\textstyle >}{\sim}\;$}}

\begin{document}



\preprint{Published in PRD as DOI:10.1103/PhysRevD.94.122007}

\title{Ultrahigh-energy neutrino follow-up of Gravitational Wave events \\
GW150914 and GW151226 with the Pierre Auger Observatory}


\author{A.~Aab}
\affiliation{Universit\"at Siegen, Fachbereich 7 Physik -- Experimentelle Teilchenphysik, Germany}

\author{P.~Abreu}
\affiliation{Laborat\'orio de Instrumenta\c{c}\~ao e F\'\i{}sica Experimental de Part\'\i{}culas -- LIP and Instituto Superior T\'ecnico -- IST, Universidade de Lisboa -- UL, Portugal}

\author{M.~Aglietta}
\affiliation{Osservatorio Astrofisico di Torino (INAF), Torino, Italy}
\affiliation{INFN, Sezione di Torino, Italy}

\author{I.~Al Samarai}
\affiliation{Laboratoire de Physique Nucl\'eaire et de Hautes Energies (LPNHE), Universit\'es Paris 6 et Paris 7, CNRS-IN2P3, France}

\author{I.F.M.~Albuquerque}
\affiliation{Universidade de S\~ao Paulo, Inst.\ de F\'\i{}sica, S\~ao Paulo, Brazil}

\author{I.~Allekotte}
\affiliation{Centro At\'omico Bariloche and Instituto Balseiro (CNEA-UNCuyo-CONICET), Argentina}

\author{A.~Almela}
\affiliation{Instituto de Tecnolog\'\i{}as en Detecci\'on y Astropart\'\i{}culas (CNEA, CONICET, UNSAM), Centro At\'omico Constituyentes, Comisi\'on Nacional de Energ\'\i{}a At\'omica, Argentina}
\affiliation{Universidad Tecnol\'ogica Nacional -- Facultad Regional Buenos Aires, Argentina}

\author{J.~Alvarez Castillo}
\affiliation{Universidad Nacional Aut\'onoma de M\'exico, M\'exico}

\author{J.~Alvarez-Mu\~niz}
\affiliation{Universidad de Santiago de Compostela, Spain}

\author{M.~Ambrosio}
\affiliation{INFN, Sezione di Napoli, Italy}

% ...

\collaboration{The Pierre Auger Collaboration}
\email{auger_spokespersons@fnal.gov}
\homepage{http://www.auger.org}
\noaffiliation

\begin{abstract}
On September 14, 2015 the Advanced LIGO detectors observed their first gravitational-wave 
(GW) transient GW150914. This was followed by a second GW event observed
on December 26, 2015. Both events were inferred to have arisen from the merger of black holes in binary systems.
Such a system may emit neutrinos if there are magnetic fields and disk debris remaining
from the formation of the two black holes. 
With the surface detector array of the Pierre Auger Observatory we can search for neutrinos 
with energy $E_\nu$ above 100 PeV from point-like sources across the sky 
with equatorial declination from about $-65^\circ$ to $+60^\circ$, and in particular from 
a fraction of the 90\% confidence-level (CL) inferred positions in the sky of GW150914 and GW151226. 
A targeted search for highly-inclined extensive air showers, produced either by interactions 
of downward-going neutrinos of all flavors in the atmosphere or by the decays of tau leptons 
originating from tau-neutrino interactions in the Earth's crust (Earth-skimming neutrinos), 
yielded no candidates in the Auger data collected within $\pm 500$ s around or 1 day after the coordinated universal time (UTC)  
of GW150914 and GW151226, as well as in the same search periods relative to the UTC time of the GW candidate event LVT151012. 
From the non-observation we constrain the amount of energy radiated in 
ultrahigh-energy neutrinos from such remarkable events.  
\end{abstract}



\date{August 4, 2016}

\maketitle

\section{Introduction}

On September 14, 2015 at 09:50:45 UTC the Advanced LIGO detectors observed the first gravitational-wave 
transient GW150914 \cite{LIGO_GW150914}. 
The GW was inferred to have arisen from the merger of black holes in a binary system
at a luminosity distance $D_s=410^{+160}_{-180}$ Mpc. 
The estimated amount of energy released in the form of 
gravitational waves was $E_{\rm GW} = 3.0^{+0.5}_{-0.5}~M_{\odot} c^2$ solar masses
\cite{LIGO_GW150914,LIGO_GW150914_Properties}.
A second GW event GW151226 \cite{LIGO_GW151226} was detected at 03:38:53 UTC on December 26, 2015, 
also inferred to be produced by the merger of two black holes at a distance $D_s=440^{+180}_{-190}$ Mpc. 
In this case the amount of energy released in the form of GW was 
$E_{\rm GW} = 1.0^{+0.1}_{-0.2}~M_{\odot} c^2$ \cite{LIGO_GW151226}.
A third candidate event, LVT151012,
was observed on October 12, 2015 at 09:54:43 UTC. 
Although LVT151012 is consistent with a binary black-hole merger
it is not significant enough to claim an unambiguous detection \cite{LIGO_3GW}.

The observation of GW events with LIGO has motivated several 
models on the production of electromagnetic counterparts 
to GW in binary black-hole mergers \cite{Bartos_BH,Stone_BH}.
Moreover, observations with the Fermi GBM detector
have revealed the presence of a transient source above 50 keV, 
only 0.4 s after GW150914, with localization consistent with its direction \cite{Fermi_GBM_GW} 
and with a possible association with a short gamma-ray burst 
\cite{Meszaros_GW_GRB,Perna_GW_GRB,Murase_GW_GRB}.
On the other hand, other gamma-ray and X-ray observatories did not find any potential counterpart
either for GW150914 \cite{INTEGRAL_GW,Fermi_LAT_GW,XMM_GW,AGILE_GW} 
or for GW151226 \cite{Fermi_LAT_GBM_GW}.
 
Mergers of black holes are a potential environment where cosmic rays can be accelerated 
to ultrahigh-energies (UHEs) provided there are magnetic fields and disk debris remaining 
from the formation of the black holes \cite{Kotera-Silk,Murase_GW_GRB}. 
These are two necessary ingredients to accelerate cosmic rays to ultrahigh energies 
through the Fermi mechanism at astrophysical sources (see for instance \cite{Gaisser_review}).
The estimated rate of this type of mergers can 
account for the total energy observed in ultrahigh-energy cosmic rays (UHECRs) 
and their distribution in the sky \cite{Kotera-Silk}. 
The UHE cosmic rays can interact with the surrounding matter or radiation 
to produce ultrahigh-energy gamma rays and neutrinos \cite{Kotera-Silk,Murase_GW_GRB}.
Other models speculate on the possibility that protons could be accelerated up to 
$\sim 10$ EeV energies in a one-shot boost \cite{Anchordoqui_GW}.
Collisions of UHE protons with photon backgrounds and gas surrounding the black hole 
would produce UHE neutrinos. The remarkable power of GW150914 could produce a proton
spectrum peaked at EeV energies with a lesser emission of neutrinos in the PeV energy range \cite{Anchordoqui_GW}. 
Neutrino experiments with peak sensitivities in the TeV-PeV energy range 
such as IceCube and ANTARES have reported no neutrino candidates 
in spatial and temporal coincidence with GW150914 \cite{GW_IceCube_ANTARES}. 

With the surface detector (SD) of the Pierre Auger Observatory \cite{Auger_detector} 
we can identify neutrino-induced 
air showers in the energy range above 100 PeV \cite{combined}. 
Showers induced by neutrinos at large zenith angles can start their development 
deep in the atmosphere so that they have a considerable amount of electromagnetic 
component at the ground (``young'' shower front).
On the other hand, at large zenith angles the atmosphere is thick enough 
that the electromagnetic component of the more numerous nucleonic cosmic rays, 
which interact shortly after entering the atmosphere, 
gets absorbed and the shower front at ground level is dominated by muons  
(``old'' shower front).  
The SD consists of $1660$ water-Cherenkov stations 
spread over an area of ${\sim} 3000~{\rm km^2}$, separated by $1.5~{\rm km}$ 
and arranged in a triangular grid. 
Although the SD is not separately sensitive 
to the muonic and electromagnetic components of the shower, 
nor to the depth at which the shower is initiated,  
the signals produced by the 
passage of shower particles, digitized with 
25 ns time resolution \cite{Auger_detector}, 
allow us to distinguish narrow traces 
in time induced by inclined showers initiated high in the atmosphere, 
from the broad signals expected in inclined showers initiated close to the ground.
Applying this simple idea, with the SD of the Pierre Auger Observatory~\cite{Auger_detector} 
we can efficiently detect inclined showers and search for two types of neutrino-induced 
showers at energies above 100 PeV:
\begin{enumerate}

\item 
Earth-skimming (ES) showers induced by tau neutrinos ($\nu_\tau$)
that travel in a slightly upward direction.
A \nutau can skim the Earth's crust and interact near the surface, 
inducing a tau lepton which escapes the Earth and decays in flight in the atmosphere,
close to the SD. Typically, only \nutau-induced showers with zenith 
angles $90^\circ < \theta < 95^\circ$ may be identified. 

\item 
Showers initiated by neutrinos of any flavor moving down at large zenith angles 
$75^\circ < \theta < 90^\circ$ with respect to the vertical 
and that interact in the atmosphere close 
to the surface-detector array through charged-current or neutral-current
interactions. These will be referred to as downward-going high zenith angle (DGH) neutrinos. 

\end{enumerate}

In previous publications \cite{combined,DGH,PRL_nutau,nu_tau_long} 
methods were established to identify inclined and deeply-initiated showers 
with the SD of the Pierre Auger Observatory.
These were applied blindly to search for ES and DGH neutrinos in
the data collected with the SD up to 20 June 2013. No neutrino candidate was found. 
As a result an upper limit to the diffuse flux of UHE neutrinos 
(i.e., from an ensemble of unresolved sources) was obtained in \cite{combined}. 
Also the same analysis was applied to place upper limits on continuous (in time) point-like
sources of UHE neutrinos \cite{Auger_PS}. 

In this paper we use the same identification criteria as in \cite{combined} to search for neutrinos in 
temporal and spatial coincidence with GW150914 and GW151226, 
as well as with the GW candidate event LVT151012 \cite{LIGO_3GW}.
The search was performed within $\pm 500$ s around
the time of either GW event as well as in the period of 1 day after their occurrence.
The choice of these two rather broad time windows is motivated by the association of mergers of 
compact systems and gamma-ray bursts (GRBs) \cite{Meszaros_GW_GRB,Perna_GW_GRB,GRB_review}. 
The $\pm 500$ s window \cite{Baret} corresponds to an 
upper limit on the duration of the prompt phase of GRBs, when typically PeV neutrinos are 
thought to be produced in interactions of accelerated cosmic rays and the gamma rays within the 
GRB itself. The choice of the 1-day window after the GW event 
is a conservative upper limit on the duration of GRB afterglows, where ultrahigh-energy 
neutrinos are thought to be produced in interactions of UHECRs with the lower-energy 
photons of the GRB afterglow (see \cite{GRB_review} for a review).

The results of the search allow us to set constraints on the emission of UHE neutrinos 
from the merger of two black holes.
These constraints apply in the energy range [$\sim$ 100 PeV, $\sim$ 25 EeV] and are complementary to those 
of IceCube/ANTARES \cite{GW_IceCube_ANTARES} which apply in the energy range [$\sim$ 100 GeV, $\sim$ 100 PeV].  

\section{Results}

The neutrino identification criteria applied to data collected with the Pierre Auger
Observatory are summarized in reference \cite{combined}.
Firstly, inclined showers are selected in the different angular 
ranges of the ES and DGH channels. Secondly, deeply-penetrating showers are identified 
in the inclined-event sample through the broad time structure of the signals expected 
to be induced in the water-Cherenkov SD stations indicative of the presence 
of an electromagnetic component \cite{combined}. 

\begin{figure}[t!]
\begin{center}$
\begin{array}{c}
\includegraphics[width=8.5cm]{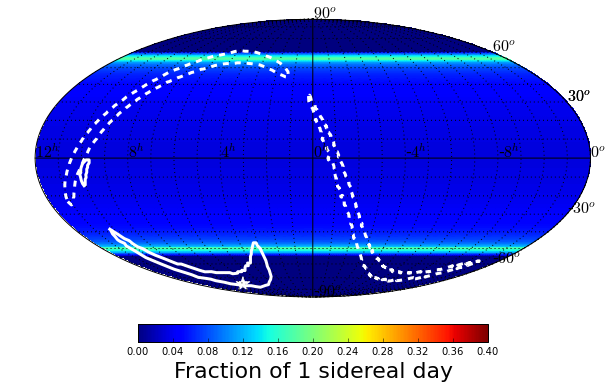} \\ 
\includegraphics[width=8.5cm]{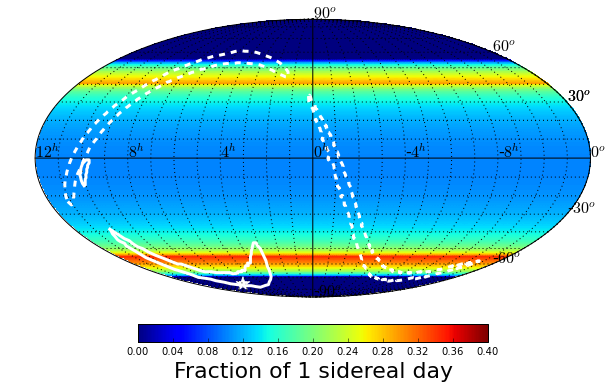}  
\end{array}$
\end{center}
\caption{
Sky map in equatorial coordinates where the color scale indicates 
the fraction of one sidereal day for which a point-like source at declination $\delta$ 
is visible to the SD of the Auger Observatory 
(latitude $\lambda=-35.2^\circ$) at zenith angle 
$90^\circ < \theta < 95^\circ$ (top panel), and 
$75^\circ < \theta < 90^\circ$ (bottom panel).
The white solid lines indicate the 90\% CL contour position of GW150914 
\cite{LIGO_GW150914,LIGO_GW150914_Properties}
and the dashed white lines indicate the corresponding 90\% CL contour position of GW151226 
\cite{LIGO_GW151226,LIGO_3GW}.
The white star indicates the best-fit 
position of the GW150914 event obtained in combination with data from 
the Fermi-GBM instrument (see Fig. 10 in \cite{Fermi_GBM_GW}).
}
\label{fig:raw-fraction}
\end{figure}

The sensitivity to UHE neutrinos in Auger is limited to large zenith angles.
As a consequence at each instant in time, neutrinos can be detected efficiently 
only from a specific portion of the sky. A source at declination $\delta$ 
and right ascension (RA) $\alpha$ (in equatorial coordinates) is seen at the latitude
of Auger ($\lambda = -35.2^\circ$) and at a given sidereal time $t$ 
with a zenith angle $\theta(t)$ given by: 
\begin{equation}
\cos\theta(t) = \sin\lambda\,\sin\delta +
\cos\lambda\,\cos\delta\,\sin(2 \pi t/T - \alpha) \,\, ,
\label{eq:costheta-t}
\end{equation}
where $T$ is the duration of one sidereal day. 
From Eq.~(\ref{eq:costheta-t}) it is straigthforward
to calculate the fraction of a sidereal day a source at declination $\delta$ is visible in 
the ES angular range $(90^\circ, ~95^\circ)$ and in the DGH one $(75^\circ, ~90^\circ)$.
In Fig.~\ref{fig:raw-fraction} we show two sky maps in equatorial 
coordinates where the color scale indicates the fraction of a sidereal day 
during which each declination is seen in the ES (top plot) and DGH (bottom plot) field of view.  
The positions of GW150914 and GW151226 are not well constrained by data collected with the 
Advanced LIGO detectors but 90\% CL contours are provided and are 
also shown in Fig.~\ref{fig:raw-fraction}. 
At 90\% CL the declination of the
source of GW150914 can be between $\delta \sim -1.0^\circ$ and $\sim -14.5^\circ$ 
or between $\delta \sim -38.5^\circ$ and $\sim -78.0^\circ$,
and that of GW151226 between $\delta \sim -72.7^\circ$ and $\sim 60.9^\circ$ 
as can be seen in Fig.~\ref{fig:raw-fraction}. Both 90\% CL declination ranges overlap with
the field of view of the ES and DGH channels for fractions of one sidereal day 
that can reach up to $\sim 17\%$ and $\sim 35\%$, respectively. 
If the emission took less time than a day these numbers could change significantly, 
depending on the sky position of the GW event relative to Auger during the emission time. 
The overlapping between the Auger field of view in the inclined directions and the 
90\% CL contour position of the GW event is larger for GW151226 
as seen in Fig.~\ref{fig:raw-fraction} and also for LVT151012.

\subsection{Searching for UHE neutrinos in coincidence with GW events}

We searched for neutrino events in coincidence with 
GW150914, GW151226 and LVT151012 in two periods of time: $\pm 500$ s around the UTC times at which 
they occurred, as well as in a period of 1 day after GW150914, GW151226 and LVT151012.

The performance of the SD array is monitored
every minute and is rather stable in each of the $\pm 500$ s and 1 day periods of 
time after either GW event. The average (root-mean squared, RMS) number of active stations 
during the search periods 
of the GW150914 and GW151226 events and of the LVT151012 candidate amount, 
respectively, to $\sim 97.5\%$ ($\sim 1.5\%$), $\sim 95.6\%$ ($\sim 5.5\%$) 
and $\sim 94.0\%$ ($6.5\%$) of the total number of stations in the SD array.  

The arrival directions of cosmic rays are determined in Auger 
from the relative arrival times of the shower front in the triggered stations.
The angular accuracy depends on the number of triggered stations, on the energy 
and on the zenith angle of the shower. Studies of cosmic ray-induced showers below 
$80^\circ$ zenith angle have revealed that the angular resolution is better than $2.5^\circ$, 
improving significantly as the number of triggered stations increases \cite{Bonifazi,Auger_inclined_rec}.
Similar results are expected for neutrino-induced showers.

\begin{figure}[t!]
\begin{center}$
\begin{array}{c}
\includegraphics[width=8.5cm]{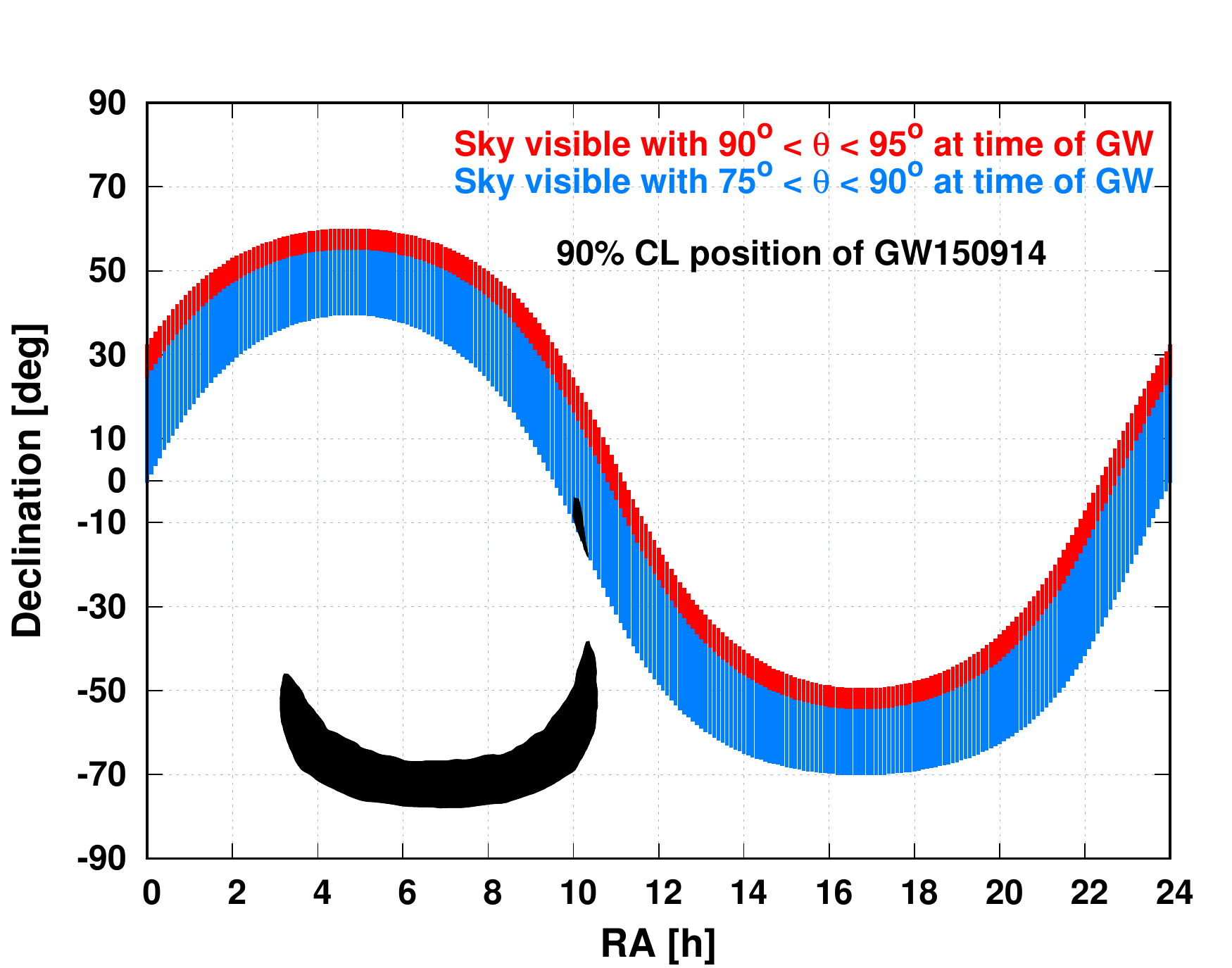} \\ 
\includegraphics[width=8.5cm]{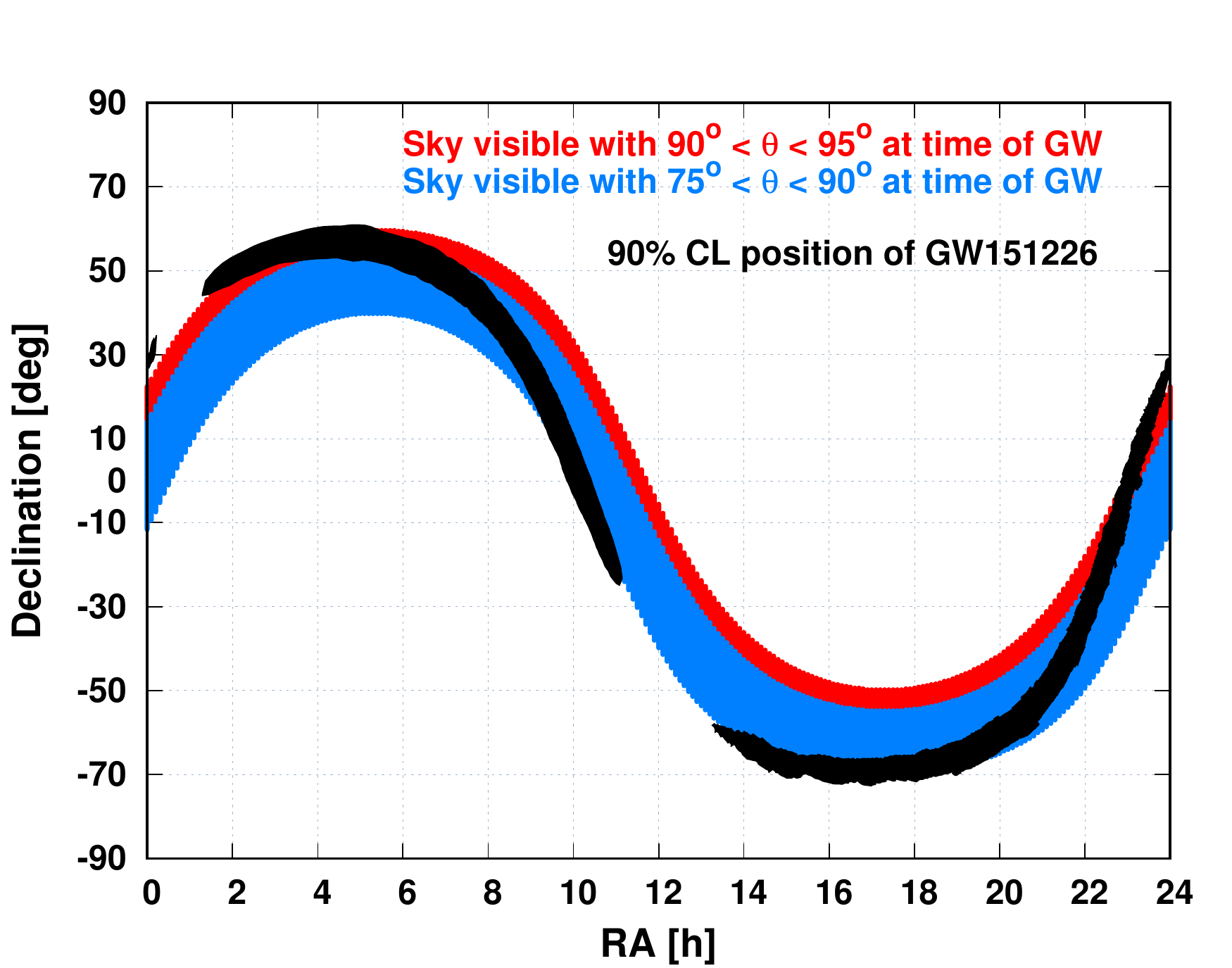}  
\end{array}$
\end{center}
\caption{
Instantaneous field of view of the ES (red band) and DGH (blue band) channels 
at the moment of coalescence of GW150914 (top panel) and of GW151226 (bottom panel).
The black spots represent the 90\% CL contour enclosing the positions
of the corresponding GW events. Note that by chance the instantaneous 
field of view of Auger is approximately the same 
at the instants of occurrence of both GW events.  
}
\label{fig:fov}
\end{figure}

Unfortunately the field of view of the ES channel did not overlap within $\pm 500$ s of the time of
coalescence of event GW150914 with the 90\% CL contour enclosing its position, 
see the top panel of Fig.~\ref{fig:fov}. 
However there is a significant overlap in the case of GW151226 as can be seen in the bottom panel 
of Fig.~\ref{fig:fov} and also in the case of LVT151012. Also GW150914,  
GW151226 and LVT151012 are visible in the DGH angular range $75^\circ < \theta < 90^\circ$ 
within $\pm 500$~s of occurrence - see Fig.~\ref{fig:fov}. 
In all cases a significant portion of the inferred position of the source 
is visible for a fraction of the time in 1 day after the corresponding GW event, 
as the Earth rotates and the field of view of the ES and DGH analyses moves 
through the sky (see Fig.~\ref{fig:raw-fraction}).

The search for UHE neutrinos in Auger data produced the following results:
\begin{itemize}

\item 
No inclined showers passing the ES or DGH selection were found 
in the time window $\pm 500$ s around GW150914 or GW151226. 

\item A total of 24 inclined showers were found with the ES  
selection criteria, $12$ in each of the 1 day periods after GW150914 and GW151226 events,  
but none of them fulfilled the neutrino identification criteria.
Also 24 and 22 inclined showers were found with the DGH selection
1 day after GW150914 and GW151226, respectively, with none of them
identified as a neutrino candidate. All selected inclined events 
have properties compatible with background nucleonic cosmic-ray events. 

\item
Also, no neutrino candidates were found within $\pm 500$ s around or 1 day after
the UTC time of the GW candidate event LVT151012 \cite{LIGO_3GW}.

\end{itemize}

\subsection{Constraints on the sources of GW}
The absence of neutrino candidates allows us to place upper limits to the UHE neutrino flux 
from GW150914 and GW151226 
(in the following we restrict ourselves to the 2 confirmed GW events)
as a function of equatorial declination $\delta$.   
The expected number of events for a neutrino flux ${dN^{\rm GW}}/{dE_\nu}(E_\nu)$ from a point-like 
source at declination $\delta$ is given by
\begin{equation}
N^{\rm GW}_{\rm event} = \int_{E_\nu}~\frac{dN^{\rm GW}_\nu}{dE_\nu}(E_\nu)~{\cal E}_{GW}(E_\nu,\delta)~dE_\nu,
\end{equation}
where ${\cal E}_{\rm GW}(E_\nu,\delta)$ is the effective exposure to a point-like flux of UHE neutrinos
as a function of neutrino energy $E_\nu$ and declination. 
For each channel ES and DGH we calculate the exposure to UHE neutrinos  
${\cal E}^{\rm ES}(E_\nu,\delta)$ and ${\cal E}^{\rm DGH}(E_\nu,\delta)$, respectively, 
following the procedure explained in \cite{combined,DGH,PRL_nutau,nu_tau_long,Auger_PS}. 
The exposure is obtained by integrating  
the SD aperture (area $\times$ solid angle) over the search period $T_{\rm search}$,
multiplied by the neutrino cross section for each neutrino channel, and weighted 
by the selection and detection efficiency obtained from Monte Carlo simulations \cite{combined}. 
When integrating over the search period, we only consider the fraction of time when the source is 
visible from the SD of Auger within the zenith angle range of the corresponding neutrino selection. 
In any of the search periods the performance of the SD array was very stable, in particular there
were no large periods of inactivity as confirmed using the continuous monitoring of the Auger SD array.

Assuming a standard $E_\nu^{-2}$ energy dependence for a constant UHE neutrino flux per flavor from the source of 
GW150914 or GW151226, namely,
${dN^{\rm GW}_\nu}/{dE_\nu} = k^{\rm GW} E_\nu^{-2}$,
a $90\%$ CL upper limit on $k^{\rm GW}$ can be obtained as
\begin{equation}
k^{\rm GW}(\delta) = \frac{2.39}{\int_{E_\nu}~E^{-2}_\nu~{\cal E}_{\rm GW}(E_\nu,\delta)~dE_\nu}.
\label{eq:k_GW}
\end{equation}
We applied Eq.~(\ref{eq:k_GW}) to obtain upper limits to the 
normalization of the flux $k^{\rm GW}_{\rm ES}(\delta)$ and $k^{\rm GW}_{\rm DGH}(\delta)$
in each channel. 
The combined upper limit to the normalization $k^{\rm GW}(\delta)$ of the flux 
is obtained as
${(k^{\rm GW})}^{-1} = {(k^{\rm GW}_{\rm ES})}^{-1} + {(k^{\rm GW}_{\rm DGH})}^{-1}$.

Systematic uncertainties are incorporated in the upper limit in Eq.~(\ref{eq:k_GW}) 
and were taken into account using a semi-Bayesian extension \cite{Conrad} of the 
Feldman and Cousins approach \cite{Feldman-Cousins} (see Table II in \cite{combined} 
for a detailed account of the main sources of systematic uncertainties).

\begin{figure}[t!]
\begin{center}$
\begin{array}{c}
\includegraphics[width=8.3cm,angle=0]{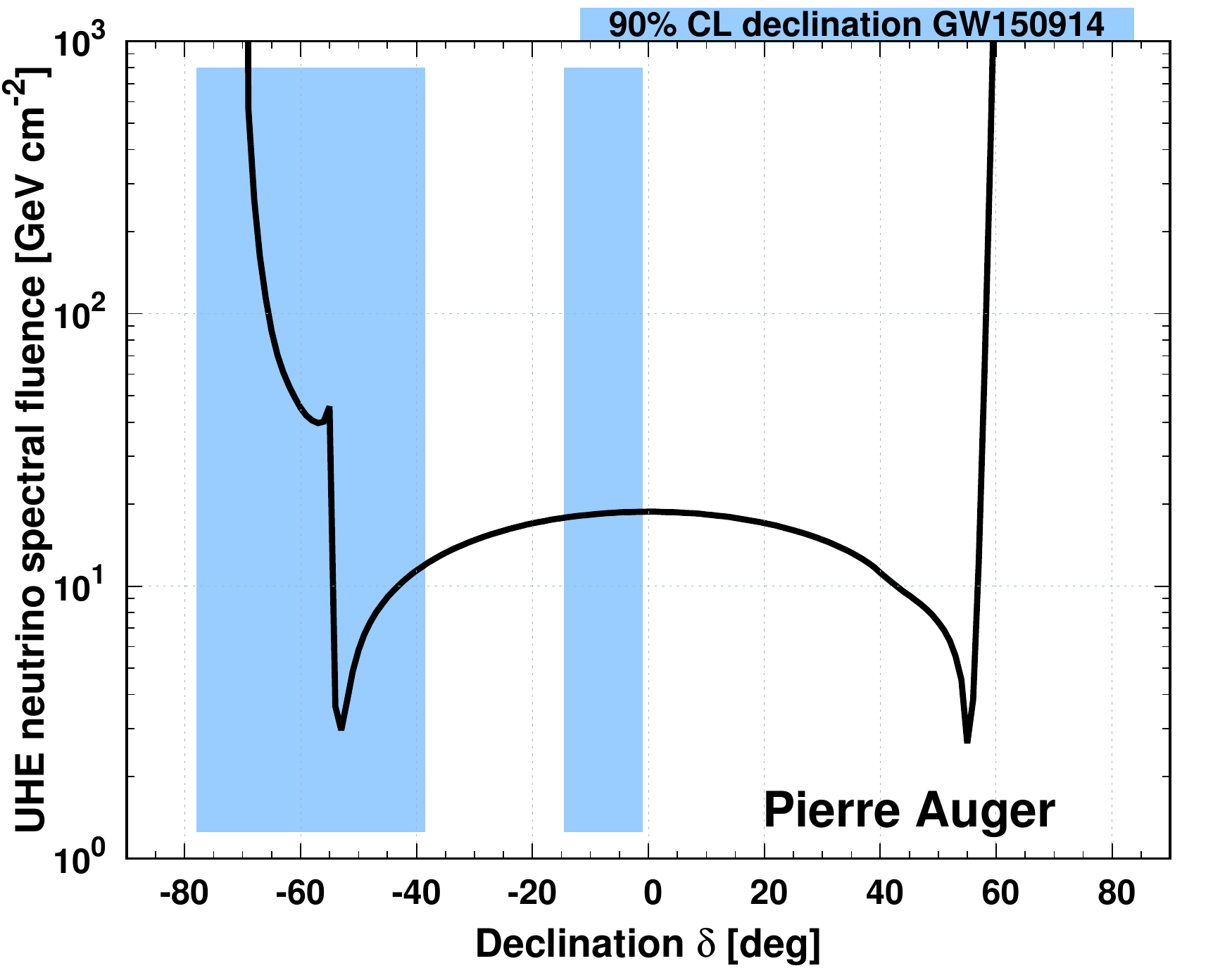} \\
\includegraphics[width=8.3cm,angle=0]{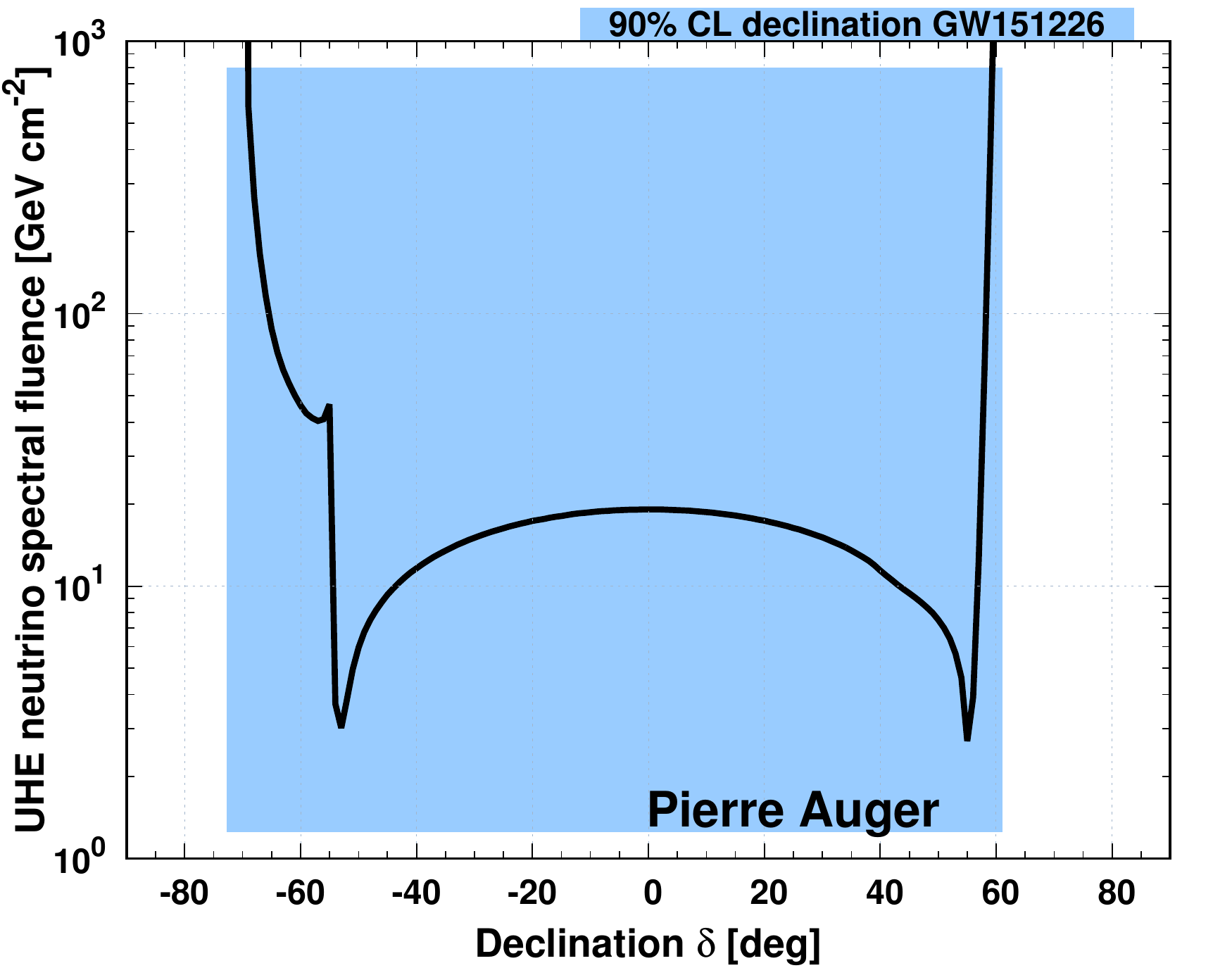}
\end{array}$
\end{center}
\caption{
Top panel: Upper limits to the UHE neutrino spectral fluence per flavor
(see Eq.~(\ref{eq:fluence})) from the source of GW150914 as a function 
of equatorial declination $\delta$. 
Fluences above the black solid line are excluded at 90\% CL
from the non-observation of UHE neutrino events in Auger.  
The 90\% CL declination bands of the GW150914 are indicated in 
the plot by the shaded rectangles.
Bottom panel: Same as the top panel for the GW event GW151226.
}
\label{fig:flux_and_fluence}
\end{figure}

From the limits to the flux normalization 
we obtained upper limits to the UHE neutrino spectral fluence radiated per flavor 
in a similar fashion to those obtained in \cite{GW_IceCube_ANTARES}:
\begin{equation}
E_\nu^2~\frac{dN_\nu}{dE_\nu} \times T_{\rm search} = k^{\rm GW}(\delta)~T_{\rm search}
\label{eq:fluence}
\end{equation}
where $T_{\rm search}= 1~ {\rm day}~+~500~{\rm s}$ is the total search period interval.
Here it is assumed that the sources of GW events emit UHE neutrinos continuously
during the search period.
The constraints on spectral fluence are shown in Fig.~\ref{fig:flux_and_fluence} and depend strongly 
on the source direction. The dependence is mainly driven by the  
fraction of the time a source at declination $\delta$ is within the  
field of view of the ES and DGH analyses. The upper limit to the fluence is 
dominated by the intrinsically-larger sensitivity of the 
ES analysis to UHE neutrinos at energies above 100 PeV. The constraints on the spectral fluence are  
above $3~{\rm GeV~cm^{-2}}$ and are very similar for both GW150914 
and GW151226 as shown in Fig.~\ref{fig:flux_and_fluence}, since the performance 
and number of active water-Cherenkov stations of the SD array are equally stable 
in each of the 1-day periods of time after each GW event.

\begin{figure}[t!]
\begin{center}$
\begin{array}{c}
\includegraphics[width=8.5cm,angle=0]{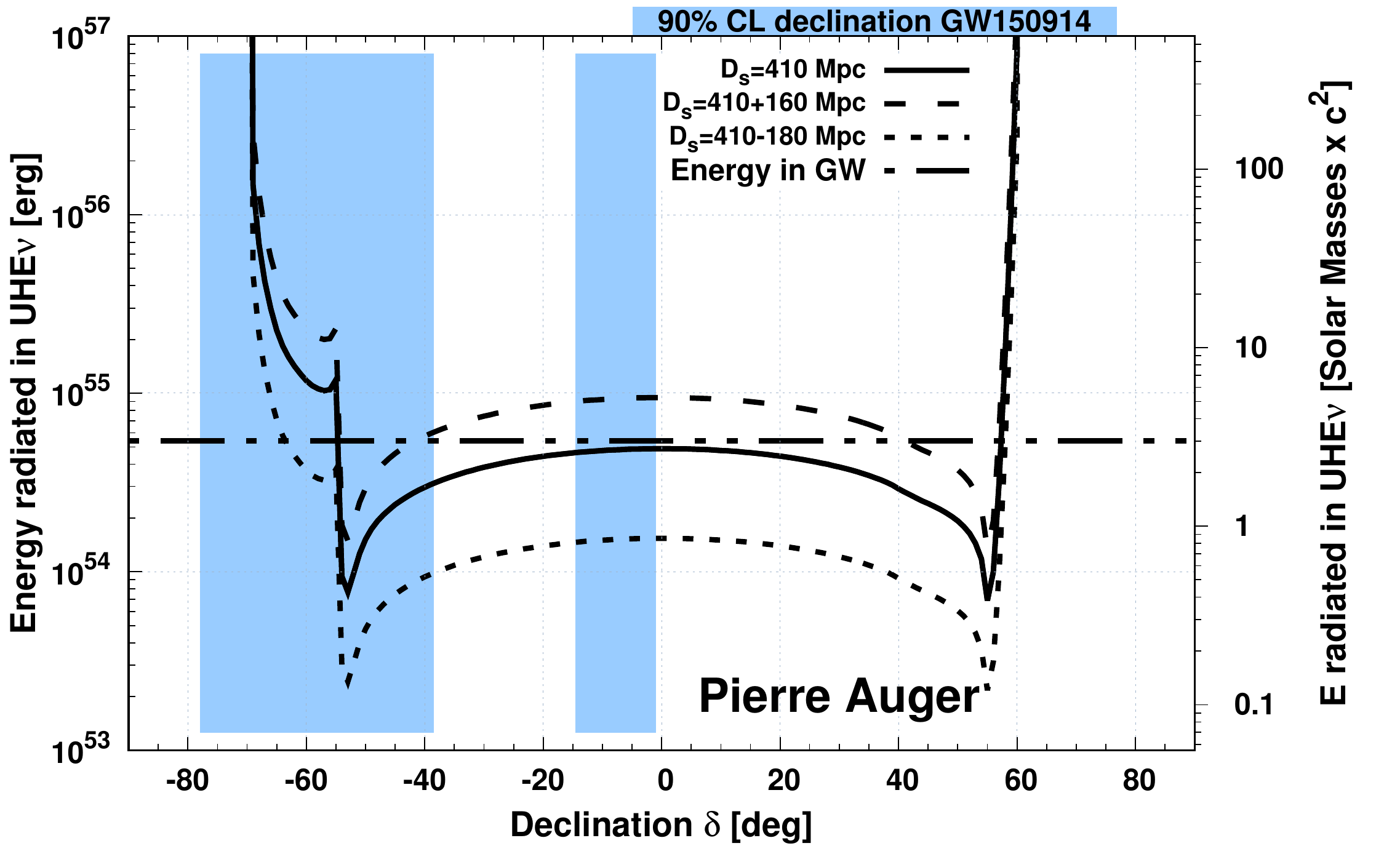} \\
\includegraphics[width=8.5cm,angle=0]{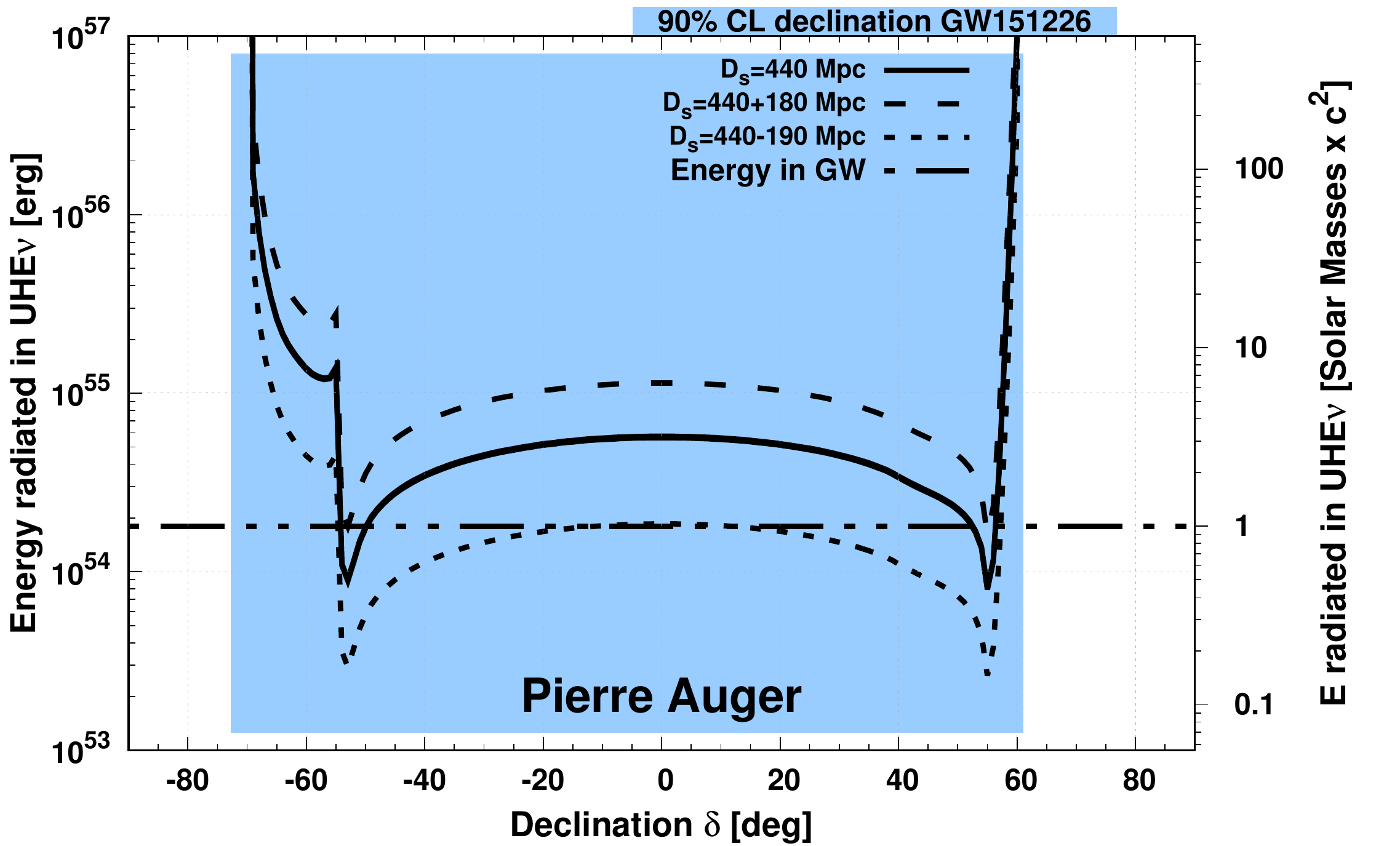}
\end{array}$
\end{center}
\caption{
Top panel: Constraints on $E_{\nu,{\rm tot}}$ the energy radiated in UHE neutrinos (per flavor) 
from the source of GW150914 as a function of equatorial declination $\delta$. 
Energies above the black solid line - assuming the luminosity distance to the source is $D_s=410$ Mpc - 
are excluded at the 90\% CL from the non-observation of UHE neutrinos in Auger. 
The long-dashed line 
represents the constraints if the source is farther away at $D_s=410+160$ Mpc, and the short-dashed line
if the source is closer to Earth at $D_s=410-180$ Mpc corresponding to the 90\% CL interval of possible 
distances to the source. For reference the dot-dashed black horizontal line represents 
$E_{\rm GW}\simeq 5.4 \times 10^{54}$ erg, the inferred energy radiated in gravitational waves 
from GW150914 \cite{LIGO_GW150914,LIGO_GW150914_Properties}.  
The limits 90\% CL declination bands of the GW150914 are indicated in 
the plot by the shaded rectangles.
Bottom panel: Same as the top panel for GW151226 but in this case $D_s=440^{+180}_{-190}$ Mpc 
and the energy released
in the form of GW is $E_{\rm GW}\simeq 1.8 \times 10^{54}$ erg.
}
\label{fig:energy}
\end{figure}

Assuming that the radiated spectrum has a 
$E_\nu^{-2}$ dependence on neutrino energy above $E_\nu=100$ PeV \cite{Gaisser_review}, 
the corresponding upper limit to the total fluence is obtained by integrating the spectral 
fluence over the interval. 
Finally, it is straightforward to obtain constraints on the total energy radiated in neutrinos 
$E_{\nu,{\rm tot}}(\delta)$ assuming the source is located at a luminosity distance $D_s$, 
$E_{\nu,{\rm tot}}(\delta) = {{\cal F}_\nu}(\delta) \times 4\pi D_s^2$.
These constraints are shown in Fig.~\ref{fig:energy}.

The most restrictive upper limits 
on the total energy emitted per flavor in UHE neutrinos
are achieved at declination $\delta\sim-53^\circ$  
\begin{equation}
E_{\nu,{\rm tot}}(\delta=-53^\circ)~<~7.7\times 10^{53}{\rm erg}, 
~~ {\rm for}~~{\rm GW150914}
\end{equation}
and at $\delta\sim 55^\circ$  
\begin{equation}
E_{\nu,{\rm tot}}(\delta=55^\circ)~<~7.9\times 10^{53}{\rm erg}, 
~~ {\rm for}~~{\rm GW151226}.
\end{equation}

The constraints on total energy can be expressed as fractions $f_\nu$ of 
energy in UHE neutrinos $E_{\nu,{\rm tot}}$ 
relative to the energy radiated in gravitational waves $E_{\rm GW}$.
The most stringent upper limit on the fraction $f_\nu$ of energy radiated in UHE neutrinos relative 
to the energy emitted in GW150914 is 
\begin{equation}
f_\nu(\delta=-53^\circ) ~<~ 14.3\% ~~ {\rm for}~~{\rm GW150914},
\end{equation}
assuming the source is located at the central value of the 
90\% CL interval of distances $D_s=410$ Mpc. 
This fraction changes from $\sim 4.5\%$ to $\sim 27.6\%$ as the source distance 
varies between the lower and upper limits of the $90\%$ CL interval $D_s=(230, 570)$ Mpc 
quoted in \cite{LIGO_GW150914}.

For the case of GW151226 since the total energy released in GW
is 3 times smaller the corresponding best upper limit on $f_\nu$ is:
\begin{equation}
f_\nu(\delta=55^\circ) ~<~ 44.1\% ~~ {\rm for}~~{\rm GW151226},
\end{equation}
assuming the source is located at $D_s=440$ Mpc. 
%

\section{Discussion}

The results in this work represent the first upper limits on UHE neutrino emission from an identified
source of GW - the merger of two black holes - and the first follow-up of GW events with neutrinos 
of energies above 100 PeV.

The upper limits on fluence emitted in the form of UHE neutrinos 
are strongly declination-dependent. 
With the SD of the Pierre Auger Observatory we are sensitive to a large fraction 
of the declination range in which the sources of GW150914 and GW151226 
could be located at the 90\% CL as shown in Fig.~\ref{fig:raw-fraction}. 


While our most stringent upper limit to the total energy in the form of UHE neutrinos 
for the GW150914 event is $\sim 7.7\times 10^{53}$ erg per flavor 
at $\delta_0=-53^\circ$, the IceCube/ANTARES best upper limit  
$(\nu_\mu+\bar\nu_\mu)$ is $\sim 5.4 \times 10^{51}$ erg 
at declinations close to the equator \cite{GW_IceCube_ANTARES}. 
However, the IceCube/ANTARES limits apply in the energy 
range [100 GeV, 100 PeV] while the Auger limits apply 
in the complementary energy range [100 PeV, 25 EeV].

In \cite{Kotera-Silk} it was argued that black-hole mergers would have sufficient 
luminosity to power the acceleration of cosmic rays up to 100 EeV.  
With a modest efficiency $\lesssim 0.03$ per GW event per unit of gravitational-wave energy 
release radiated in the form of UHECRs and given the inferred rate of BH mergers \cite{LIGO_3GW}, 
a source population of this type could achieve the energy budget needed to explain 
the observed UHECRs \cite{Kotera-Silk}. In this work we place a most stringent upper limit 
on the fraction of GW energy channeled into neutrinos of $\sim 14\%$. 
If only $3\%$ of the energy of the GW is channeled into UHECRs \cite{Kotera-Silk}, 
and the same energy goes into UHE neutrinos, then we would expect at most on the order of 
$0.5$ events in Auger in coincidence with GW150914.

An upper bound to the diffuse single-flavor neutrino flux integrated over 
a source population of this type was estimated also in \cite{Kotera-Silk},
\begin{equation}
E_\nu^2~\frac{dN_\nu}{dE_\nu}\biggr \rvert^{\rm theory}_{\rm diffuse} 
\lesssim (1.5 - 6.9) \times 10^{-8}~{\rm GeV~cm^{-2}~s^{-1}~sr^{-1}},
\end{equation}
depending on the evolution with redshift of the sources and assuming an optical depth $\tau = 1$ 
to neutrino production in the debris surrounding the BH mergers. 
This upper bound is a factor between $\sim 3$ and 10 above the limit to the diffuse flux of 
UHE neutrinos obtained with Auger data up to 20 June 2013 in \cite{combined}, namely,
\begin{equation}
E_\nu^2~\frac{dN_\nu}{dE_\nu}\biggr \rvert^{\rm Auger}_{\rm diffuse} 
< 6.4 \times 10^{-9}~{\rm GeV~cm^{-2}~s^{-1}~sr^{-1}}.
\end{equation}
It is possible that there are no significant fluxes of UHE neutrinos associated 
with the coalescence of black holes, more phenomenological work in this area is needed.
In the case that cosmic rays are indeed 
accelerated as suggested in \cite{Kotera-Silk}, our constraints on the diffuse flux 
of UHE neutrinos would imply that either
(1) the optical depth to neutrino production is significantly smaller than 1 as 
expected in GRB models; or (2) only a fraction of the luminosity that can be extracted from 
the BH can be invested in UHECRs acceleration, or (3) only a fraction of the energy of 
the protons goes into charged pions (that are the parents of the neutrinos); 
or (4) a combination of the three possibilities. 

The Advanced LIGO-Virgo detection of GW150914 and GW151226 represents a breakthrough in 
our understanding of the Universe. Similar analyses to those presented in this work 
will be important to provide constraints on the progenitors of the GW emission. 
Given the inferred rate of events $9 - 240 ~{\rm Gpc^{-3} yr^{-1}}$  \cite{LIGO_3GW}  
new GW events can be expected in the near future, closer to Earth and/or more energetic, 
and/or produced by another type of source that is more likely to accelerate UHECRs and produce UHE neutrinos 
than the merger of two black holes, such as for instance binary neutron-star mergers, 
and core-collapse supernovae with rapidly-rotating cores \cite{NS_GW,GW_review}. 

Finally, the detection of UHE neutrino candidates in Auger in 
coincidence with GW events could help in pinpointing the position of the 
source of GW with an accuracy that depends on the shower zenith angle and energy, 
ranging from less than $\sim 1~{\rm deg}^2$ to order $10~{\rm deg}^2$ in the least favourable cases.
This is to be compared with the currently known position of the two GW events, namely 
a few $100~{\rm deg}^2$. Observations with Auger can significantly constrain the position of the source
and help the follow-up of the GW events with optical and other observatories of 
electromagnetic radiation. This is an example where multimessenger observations 
(GW, neutrinos and photons) can reveal properties of the sources which may not 
be discerned from one type of signal alone.

\section*{Acknowledgments}

\begin{sloppypar}
The successful installation, commissioning, and operation of the Pierre Auger Observatory would 
not have been possible without the strong commitment and effort from the technical and administrative 
staff in Malarg\"ue. We are very grateful to the following agencies and organizations for financial support:
\end{sloppypar}

\begin{sloppypar}
Comisi\'on Nacional de Energ\'\i{}a At\'omica, Agencia Nacional de Promoci\'on Cient\'\i{}fica y Tecnol\'ogica (ANPCyT), 
Consejo Nacional de Investigaciones Cient\'\i{}ficas y T\'ecnicas (CONICET), Gobierno de la Provincia de Mendoza, 
Municipalidad de Malarg\"ue, NDM Holdings and Valle Las Le\~nas, in gratitude for their continuing cooperation
 over land access, Argentina; the Australian Research Council; Conselho Nacional de Desenvolvimento Cient\'\i{}fico 
e Tecnol\'ogico (CNPq), Financiadora de Estudos e Projetos (FINEP), Funda\c{c}\~ao de Amparo \`a Pesquisa do Estado 
de Rio de Janeiro (FAPERJ), S\~ao Paulo Research Foundation (FAPESP) Grants No.\ 2010/07359-6 and No.\ 1999/05404-3, 
Minist\'erio de Ci\^encia e Tecnologia (MCT), Brazil; Grant No.\ MSMT CR LG15014, LO1305 and LM2015038 and the Czech 
Science Foundation Grant No.\ 14-17501S, Czech Republic; Centre de Calcul IN2P3/CNRS, Centre National de la Recherche 
Scientifique (CNRS), Conseil R\'egional Ile-de-France, D\'epartement Physique Nucl\'eaire et Corpusculaire (PNC-IN2P3/CNRS), 
D\'epartement Sciences de l'Univers (SDU-INSU/CNRS), Institut Lagrange de Paris (ILP) Grant No.\ LABEX ANR-10-LABX-63, 
within the Investissements d'Avenir Programme Grant No.\ ANR-11-IDEX-0004-02, France; Bundesministerium f\"ur Bildung 
und Forschung (BMBF), Deutsche Forschungsgemeinschaft (DFG), Finanzministerium Baden-W\"urttemberg, Helmholtz Alliance 
for Astroparticle Physics (HAP), Helmholtz-Gemeinschaft Deutscher Forschungszentren (HGF), Ministerium f\"ur Wissenschaft 
und Forschung, Nordrhein Westfalen, Ministerium f\"ur Wissenschaft, Forschung und Kunst, Baden-W\"urttemberg, Germany; 
Istituto Nazionale di Fisica Nucleare (INFN),Istituto Nazionale di Astrofisica (INAF), Ministero dell'Istruzione, 
dell'Universit\'a e della Ricerca (MIUR), Gran Sasso Center for Astroparticle Physics (CFA), CETEMPS Center of Excellence, 
Ministero degli Affari Esteri (MAE), Italy; Consejo Nacional de Ciencia y Tecnolog\'\i{}a (CONACYT) No.\ 167733, Mexico; 
Universidad Nacional Aut\'onoma de M\'exico (UNAM), PAPIIT DGAPA-UNAM, Mexico; Ministerie van Onderwijs, Cultuur en Wetenschap, 
Nederlandse Organisatie voor Wetenschappelijk Onderzoek (NWO), Stichting voor Fundamenteel Onderzoek der Materie (FOM), 
Netherlands; National Centre for Research and Development, Grants No.\ ERA-NET-ASPERA/01/11 and No.\ ERA-NET-ASPERA/02/11, 
National Science Centre, Grants No.\ 2013/08/M/ST9/00322, No.\ 2013/08/M/ST9/00728 and No.\ HARMONIA 5 -- 2013/10/M/ST9/00062, 
Poland; Portuguese national funds and FEDER funds within Programa Operacional Factores de Competitividade through Funda\c{c}\~ao 
para a Ci\^encia e a Tecnologia (COMPETE), Portugal; Romanian Authority for Scientific Research ANCS, CNDI-UEFISCDI partnership 
projects Grants No.\ 20/2012 and No.194/2012 and PN 16 42 01 02; Slovenian Research Agency, Slovenia; Comunidad de Madrid, 
Fondo Europeo de Desarrollo Regional (FEDER) funds, Ministerio de Econom\'\i{}a y Competitividad, Xunta de Galicia, European 
Community 7th Framework Program, Grant No.\ FP7-PEOPLE-2012-IEF-328826, Spain; Science and Technology Facilities Council, 
United Kingdom; Department of Energy, Contracts No.\ DE-AC02-07CH11359, No.\ DE-FR02-04ER41300, No.\ DE-FG02-99ER41107 
and No.\ DE-SC0011689, National Science Foundation, Grant No.\ 0450696, The Grainger Foundation, USA; NAFOSTED, Vietnam; 
Marie Curie-IRSES/EPLANET, European Particle Physics Latin American Network, European Union 7th Framework Program, 
Grant No.\ PIRSES-2009-GA-246806; and UNESCO.
\end{sloppypar}


\end{document}